\pdfoutput=1

\documentclass[11pt,twoside,a4paper,cmspaper,final,collab]{cms-tdr}
\def\svnVersion{42127M}\def\cmsCernNo{2011-003}\def\cmsCernDate{\today}\def\cmsMessage{Submitted to Physical Review Letters}

\begin{document}\cmsNoteHeader{HIN-10-003}

\hyphenation{had-ron-i-za-tion}
\hyphenation{cal-or-i-me-ter}
\hyphenation{de-vices}
\RCS$Revision: 42120 $
\RCS$HeadURL: svn+ssh://alverson@svn.cern.ch/reps/tdr2/papers/HIN-10-003/trunk/HIN-10-003.tex $
\RCS$Id: HIN-10-003.tex 42120 2011-02-26 16:18:52Z raphael $
\cmsNoteHeader{HIN-10-003} 
\title{Study of \cPZ\ boson production in PbPb collisions at $\sqrt{s_{\rm NN}} = 2.76$~TeV}
\address[cern]{CERN}
\author[cern]{The CMS Collaboration}
\date{\today}

\hyphenation{mo-di-fi-ca-tions}
\hyphenation{si-mi-lar-ly}
\hyphenation{mo-del-ling}
\hyphenation{Chro-mo-dy-na-mics}
\hyphenation{ideally}
\hyphenation{ener-gy}
\hyphenation{equi-va-lent}
\hyphenation{ele-men-ta-ry}
\hyphenation{cen-tra-li-ty}
\hyphenation{Re-la-ti-vis-tic}
\hyphenation{exis-ten-ce}
\hyphenation{si-mu-la-ted}
\hyphenation{cor-res-pon-ding}

\abstract{
A search for \cPZ\ bosons in the $\mu^+ \mu^-$ decay channel has been performed in PbPb collisions at $\sqrt{s_{\rm NN}} = 2.76$~TeV with the CMS detector at the LHC, in a 7.2~$\mu$b$^{-1}$ data sample. The number of opposite-sign muon pairs observed in the 60--120~GeV/$c^2$ invariant mass range is 39, corresponding to a yield per unit of rapidity ($y$) and per minimum bias event of $(33.8 \pm 5.5 ({\rm stat}) \pm 4.4 ({\rm syst})) \times 10^{-8}$, in the $|y|<2.0$ range. Rapidity, transverse momentum, and centrality dependencies are also measured. The results agree with next-to-leading order QCD calculations, scaled by the number of incoherent nucleon-nucleon collisions.
}

\hypersetup{%
pdfauthor={CMS Collaboration},%
pdftitle={Study of Z boson production in PbPb collisions at nucleon-nucleon center of mass energy = 2.76 TeV},%
pdfsubject={CMS},%
pdfkeywords={CMS, physics, heavy-ion, Z boson}}

\maketitle 

The hot and dense matter produced in heavy-ion collisions, often referred to as the quark-gluon plasma (QGP), can be studied in various ways. One approach is to compare measurements made in heavy-ion (AA) collisions to those in proton-proton (\Pp\Pp) and proton- \mbox{(or deuteron-)} nucleus collisions. Another way is to compare in the same AA sample the yields of particles that are modified by the QGP to those of unmodified reference particles. At the Relativistic Heavy Ion Collider (RHIC), direct photons play the reference role~\cite{Adler:2005ig}, although their measurement is complicated by copious background from $\pi^0$ and other decays, and by the existence of a parton fragmentation component which is potentially modified by the medium~\cite{Arleo:2006xb}. At the Large Hadron Collider (LHC) energies, a new and cleaner reference becomes available: the \cPZ\ boson, decaying into leptons~\cite{Kartvelishvili:1995fr,ConesadelValle:2009vp}.

Electroweak boson production is an important benchmark process at hadron colliders. At 7~TeV centre-of-mass energy, measurements in \Pp\Pp\ collisions at the LHC~\cite{Aad:2010yt,CMS:2010xn} are well described by calculations based on higher-order perturbative Quantum Chromodynamics (pQCD), using recent parton distribution functions (PDFs). In AA collisions, \cPZ\ boson production can be affected by various initial-state effects, though predictions indicate that these contributions are rather small~\cite{Kartvelishvili:1995fr,Vogt:2000hp,Zhang:2002yz,Paukkunen:2010qg,Neufeld:2010dz}.
Firstly, the mix of protons and neutrons in AA collisions (the so-called isospin effect) is estimated to modify the \cPZ\ yield by less than 3\% compared to \Pp\Pp\ collisions~\cite{Paukkunen:2010qg}. Secondly, energy loss and multiple scattering of the initial partons can also alter the \cPZ\ production, by about 3\%~\cite{Neufeld:2010dz}. The PDFs however are modified in nuclei and a depletion (shadowing) is expected for \cPZ\ bosons at the LHC, modifying their yield by as much as 20\%~\cite{Paukkunen:2010qg}. Precise measurements of \cPZ\ production in heavy-ion collisions can therefore help to constrain nuclear PDFs.

Once produced, \cPZ\ bosons decay within the medium, with a lifetime of $0.1$~fm/$c$. Their leptonic decays are of particular interest since leptons pass freely through the produced medium regardless of its nature (partonic or hadronic) and properties. Dileptons from \cPZ\ bosons can thus serve as a reference to the processes expected to be heavily modified in the QGP, such as quarkonia production, or the production of an opposite-side jet in \cPZ+jet processes~\cite{Kartvelishvili:1995fr,Neufeld:2010fj}. The \cPZ\ bosons are therefore ideally suited to serve as a  standard candle of the initial state in PbPb collisions at the LHC energies.

During the first PbPb LHC run at the end of 2010, at a centre-of-mass energy per nucleon pair of $\sqrt{s_{\rm NN}}=2.76$~TeV, \cPZ\ bosons were observed by the Compact Muon Solenoid (CMS) experiment. The measurement reported in this letter is performed with a 55 million minimum bias (MB) event sample, corresponding to an integrated luminosity of 7.2~$\mu$b$^{-1}$.

A detailed description of the CMS detector can be found in~\cite{CMS:2008zzk}. Its central feature is a superconducting solenoid of 6~m internal diameter, providing a magnetic field of 3.8~T. Within the field volume are the silicon pixel and strip tracker, the crystal electromagnetic calorimeter, and the brass/scintillator hadron calorimeter. Muons are measured in gas-ionisation detectors embedded in the steel return yoke. In addition, CMS has extensive forward calorimetry, in particular two steel/quartz-fiber \v{C}erenkov, hadron forward (HF) calorimeters, which cover the pseudorapidity range $2.9 < |\eta| < 5.2$.

In this analysis, \cPZ\ bosons are measured through their dimuon decays. The silicon pixel and strip tracker measures charged particle trajectories in the  range $|\eta| < 2.5$. It consists of 66M pixel and 10M strip detector channels. It provides an impact parameter resolution of $\sim$\,15~$\mu$m in the transverse plane. Muons are detected in the $|\eta| < 2.4$ range, with detection planes based on three technologies: drift tubes, cathode strip chambers, and resistive plate chambers. A matching of the muons to the tracks measured in the silicon tracker results in a $p_{\rm T}$ resolution between 1 and 2\%, for $p_{\rm T}$ values up to 100~GeV/$c$.

The centrality of AA collisions, i.e.~the geometrical overlap of the incoming nuclei, is related to the energy released in the collisions. In CMS, centrality is defined as percentiles of the distribution of the energy deposited in the HFs~\cite{HIN-10-004,D'Enterria:2007xr}. The centrality classes used in this analysis are 30--100\%, 10--30\% and 0--10\% (most central), ordered from the lowest to the highest HF energy deposit.

Events are preselected if they contain a reconstructed primary vertex made of at least two tracks, and an offline coincidence of both of the HFs with a total deposited energy of at least 3~GeV. These criteria reduce contributions from single-beam interactions with the environment (e.g. beam-gas and beam halo collisions with the beam pipe), ultra-peripheral electromagnetic collisions, and cosmic-ray muons. The acceptance of this selection is $(97 \pm 3)$\% of the hadronic inelastic cross section~\cite{HIN-10-004}.

The events are also selected by the two-level trigger of CMS. At the first hardware level, two muon candidates in the muon detectors are required. At the software-based higher-level, two reconstructed tracks in the muon detectors are required, each with a $p_{\rm T}$ of at least 3~GeV/$c$. In order to study the dimuon trigger efficiency, events are also collected with a single-muon trigger, requiring $p_{\rm T} > 20$~GeV/$c$. For \cPZ\ bosons, the trigger efficiency is estimated to be $\simeq 94$\%.

Muon offline reconstruction is seeded with $\simeq 99$\% efficiency by tracks in the muon detectors, called stand-alone muons. These tracks are then matched to tracks reconstructed in the silicon tracker by means of an algorithm optimised for the heavy-ion environment~\cite{Roland:2006kz,D'Enterria:2007xr}. For muons from \cPZ\ decays, the tracking efficiency is $\simeq$~85\%, less than in the \Pp\Pp\ case, as the track reconstruction requires more pixel hits to lower the number of combinations, due to the high multiplicity. Global fits of the muon and tracker tracks, called global muons, are used to obtain the results presented in this letter.

\begin{figure}[hbtp]
  \begin{center}
    \includegraphics[width=0.9\linewidth]{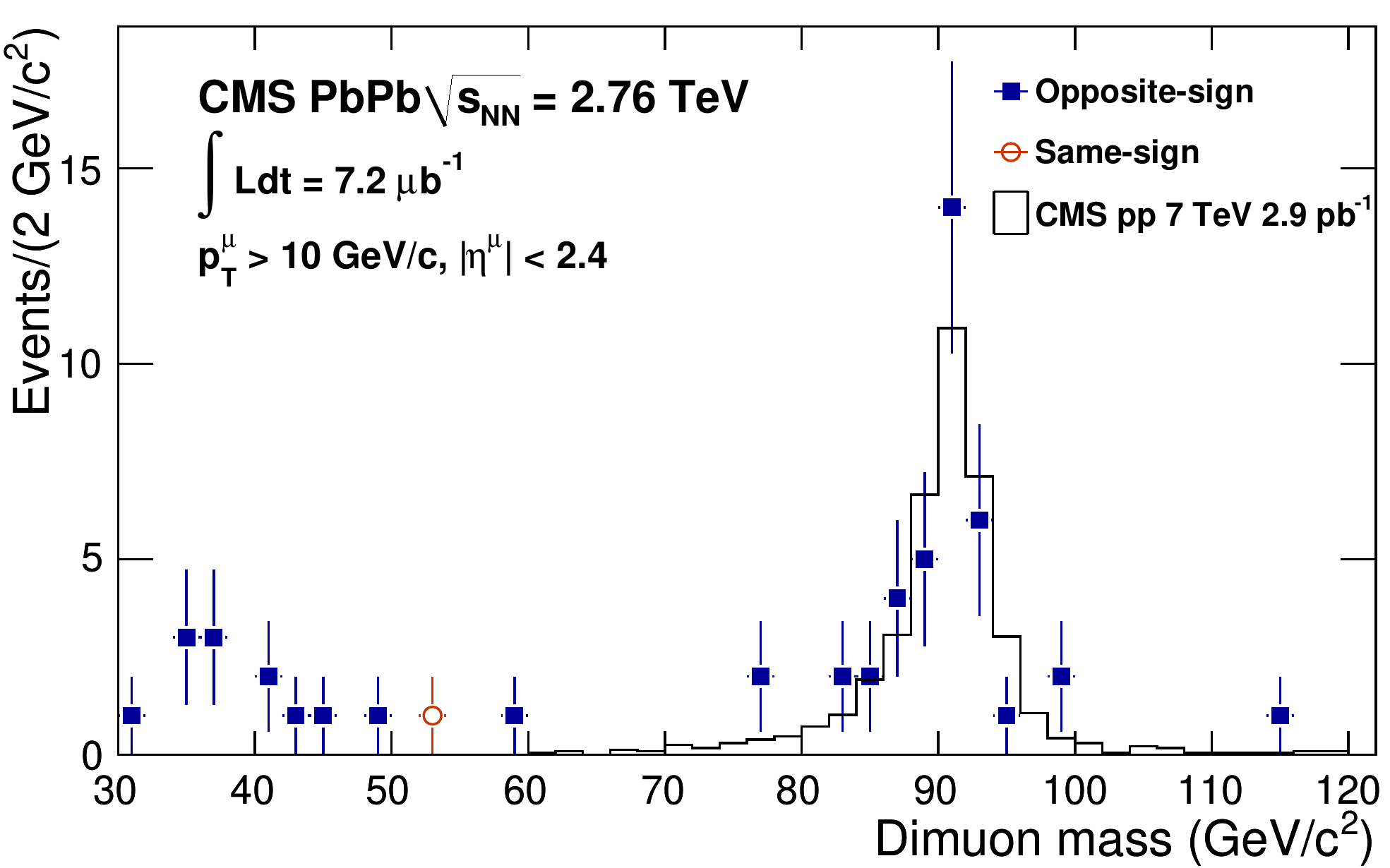} 
    \caption{Dimuon invariant mass spectra. Full squares are opposite-sign dimuons, while the empty circle shows a unique like-sign dimuon candidate. The histogram shows the corresponding distribution measured in pp collisions at 7~TeV within 60--120~GeV/$c^2$, scaled to the 39 PbPb candidates.}
    \label{fig:mass}
  \end{center}
    \vspace{-2ex} 
\end{figure}

Background muons from cosmic rays and heavy-quark semileptonic decays are rejected by requiring a transverse (longitudinal) impact parameter of less than 0.3 (1.5)~mm from the measured vertex. Loose criteria applied on the reconstructed muons result in the dimuon mass spectrum shown in Fig.~\ref{fig:mass}, for muons with $p_{\rm T} > 10$~GeV/$c$. No muon isolation criteria are applied, as they are expected to have reduced efficiency in the high particle density of the PbPb environment. The fraction of \cPZ\ decays removed by the applied selection criteria is estimated to be $\simeq 2.6$\%, in both data and simulation. Residual background is estimated to be less than 4\% by extrapolations from the low mass region, and no correction is applied. Thirty-nine \cPZ\ candidates are observed in the mass interval 60--120~GeV/$c^2$. Their distribution is consistent with the one from \Pp\Pp\ data at 7~TeV~\cite{CMS:2010xn}, scaled down to 39 counts and displayed as an histogram limited to the 60--120~GeV/$c^2$ mass range in Fig.~\ref{fig:mass}.

Muon trigger, reconstruction and selection efficiencies, as well as acceptance, are estimated using the \PYTHIA 6.424 simulation~\cite{Ref:Pythia} with CTEQ6L PDFs~\cite{Pumplin:2002vw} and full \GEANTfour~\cite{Agostinelli:2002hh} detector simulation. To take into account the effect of the higher PbPb underlying-event activity, simulated \cPZ\ decays are embedded in measured PbPb events at the level of detector hits and with generated vertices matched to the measured ones. These events were processed through the trigger emulation and event reconstruction chain. Track characteristics, such as the number of hits and the $\chi^2$ of the track fit, have similar distributions in data and simulation. The detector acceptance $\alpha$, defined as the fraction of \cPZ\ bosons produced at rapidity $|y|<2.0$ that decay into muons with $|\eta|< 2.4$ and $p_{\rm T} > 10$~GeV/$c$, is estimated to be 78\%. Within this acceptance, the overall trigger, reconstruction, and identification efficiency $\varepsilon$ averages to 67\%, and varies by less than 10\% as a function of centrality.

The individual components of this efficiency are also estimated with a technique using data, called tag-and-probe, similar to the one used for the corresponding \Pp\Pp\ measurement~\cite{CMS:2010xn}. It consists in counting the \cPZ\ candidates with and without applying the probed selection on one of the muons: 1) the stand-alone muon reconstruction efficiency is probed with tracker tracks; 2) the silicon tracker reconstruction efficiency is probed with stand-alone muons; 3) the trigger efficiency is probed by testing the trigger response to global muons from a sample triggered by a single-muon requirement. The latter is also checked with high-quality reconstructed muons from MB events. In all cases, these data-driven efficiencies agree with those derived from simulation within the statistical uncertainties.

The total systematic uncertainty on the \cPZ\ yield is estimated to be 13\% by summing in quadrature the following contributions. The largest one is associated with the tracking efficiency and taken as the 9.8\% precision of the above-mentioned data-driven efficiency determination. Similarly, the uncertainty associated with the dimuon trigger is 4.5\%. The 4\% maximum contribution from unsubtracted background is taken as a systematic uncertainty. The uncertainty associated with the selection criteria is considered to be equal to the 2.6\% loss of events. The MB trigger efficiency is known at the 3\% level. The uncertainty coming from the acceptance correction is estimated to be less than 3\%, by varying the underlying generated kinematics ($y$, $p_{\rm T}$) beyond reasonable modifications. Other systematic uncertainties are estimated to sum to less than 1.5\%.

The yield of $\cPZ \rightarrow \mu^+ \mu^-$ decays per MB event is defined as $dN/dy (|y|<2.0) = N_{\cPZ} / (\alpha \varepsilon N_{\rm MB} \Delta y )$, where $N_{\cPZ} = 39$ is the number of dimuons counted in the mass window of 60--120~GeV/$c^2$, $N_{\rm MB} = 55 \times 10^{6}$ is the number of corresponding MB events, corrected for trigger efficiency, $\alpha$ and $\varepsilon$ are the acceptance and overall efficiency, and $\Delta y = 4.0$ is the rapidity bin width. We find $dN/dy (|y|<2.0) = (33.8 \pm 5.5 \pm 4.4) \times 10^{-8}$, where the first uncertainty is statistical and the second systematic. The analysis described above is repeated after subdividing the data into three bins for each of the following variables: event centrality and \cPZ\ boson $y$ and $p_{\rm T}$. The total systematic uncertainty does not vary significantly with these variables and is considered to be constant and dominantly uncorrelated.

In the absence of in-medium modifications, the yield of perturbative processes such as the \cPZ\ boson production is supposed to scale with the number of incoherent nucleon-nucleon binary collisions. In order to compare the PbPb measured yields to available \Pp\Pp\ cross-section calculations, a scaling factor $T_{\rm AB}$ is necessary. This nuclear overlap function is equal to the number of elementary nucleon-nucleon binary collisions divided by the elementary NN cross section, and can be interpreted as the NN equivalent integrated luminosity per AA collision, at a given centrality. In units of mb$^{-1}$, the average $T_{\rm AB}$ amounts to $1.45 \pm 0.18$, $11.6 \pm 0.7$, and $23.2 \pm 1.0$, for the centrality ranges 30--100\%, 10--30\% and 0--10\%, respectively, and $5.66 \pm 0.35$ for MB events. These numbers are computed with a Glauber model calculation~\cite{Miller:2007ri}, using the same parameters as in~\cite{HIN-10-004}. The quoted uncertainties are derived by varying within uncertainties the Glauber parameters and the MB trigger and selection efficiency.

\begin{figure}[hbtp]
  \begin{center}
\includegraphics[width=0.40\textwidth]{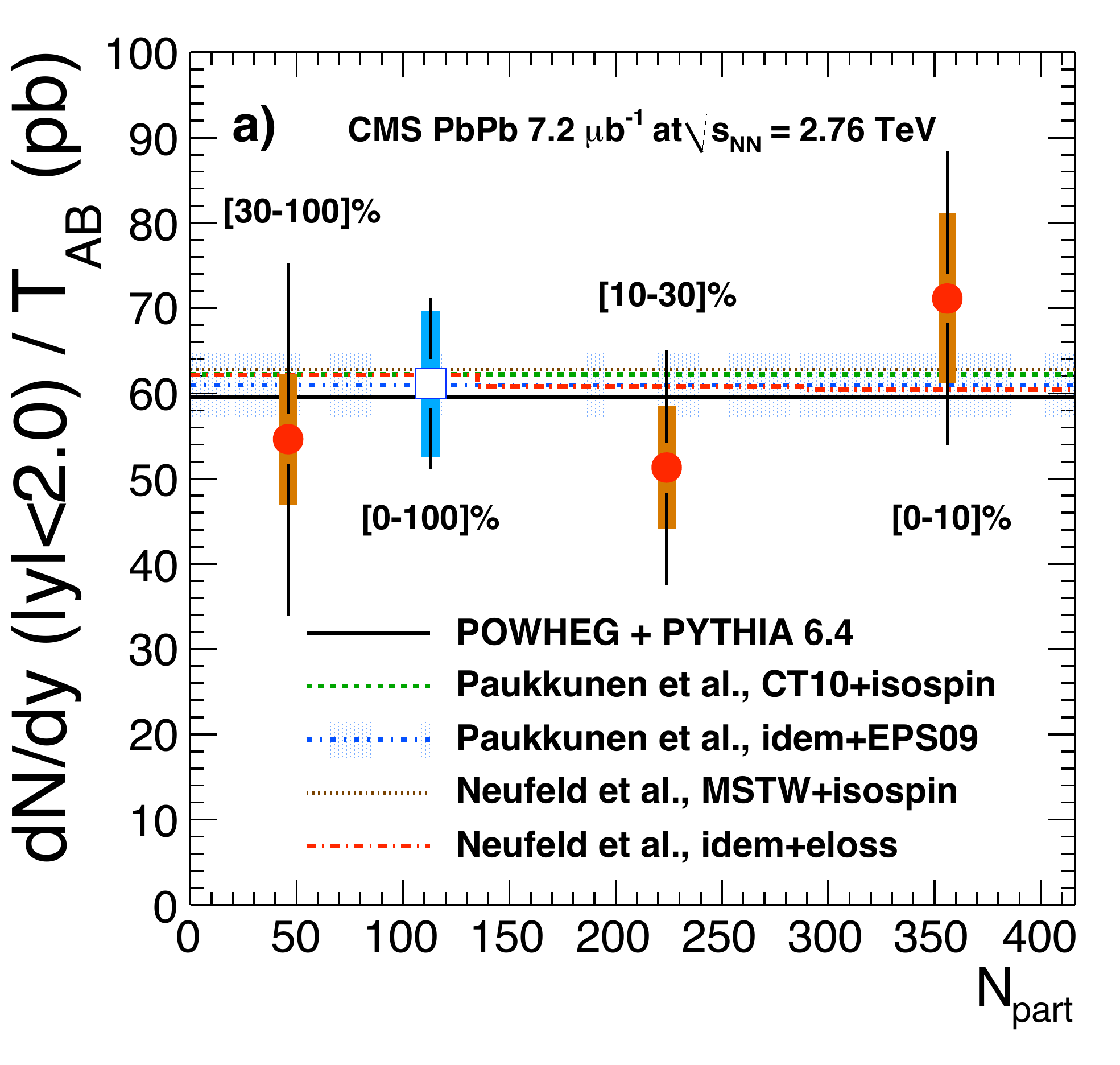}
    \includegraphics[width=0.40\textwidth]{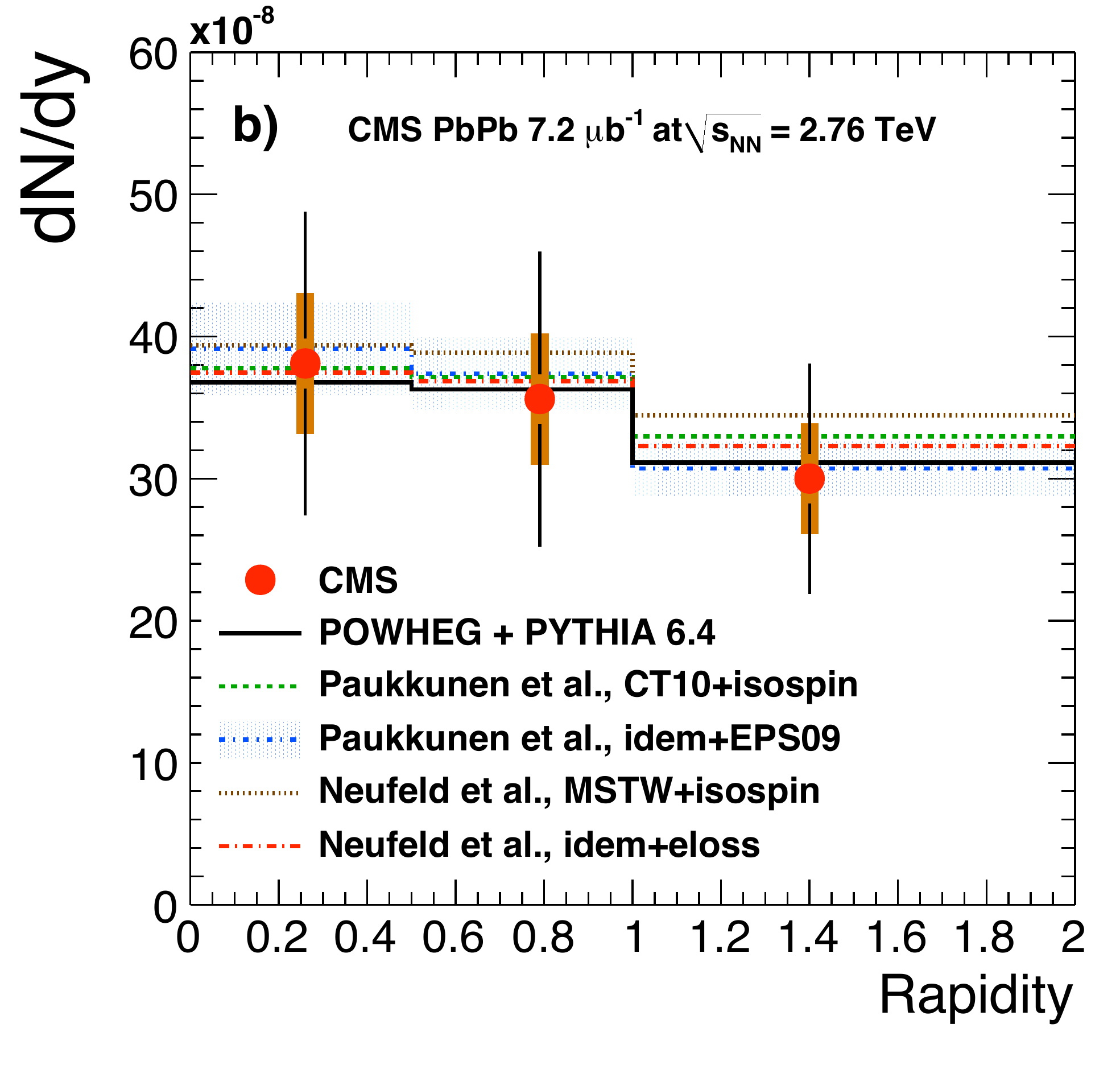} \includegraphics[width=0.40\textwidth]{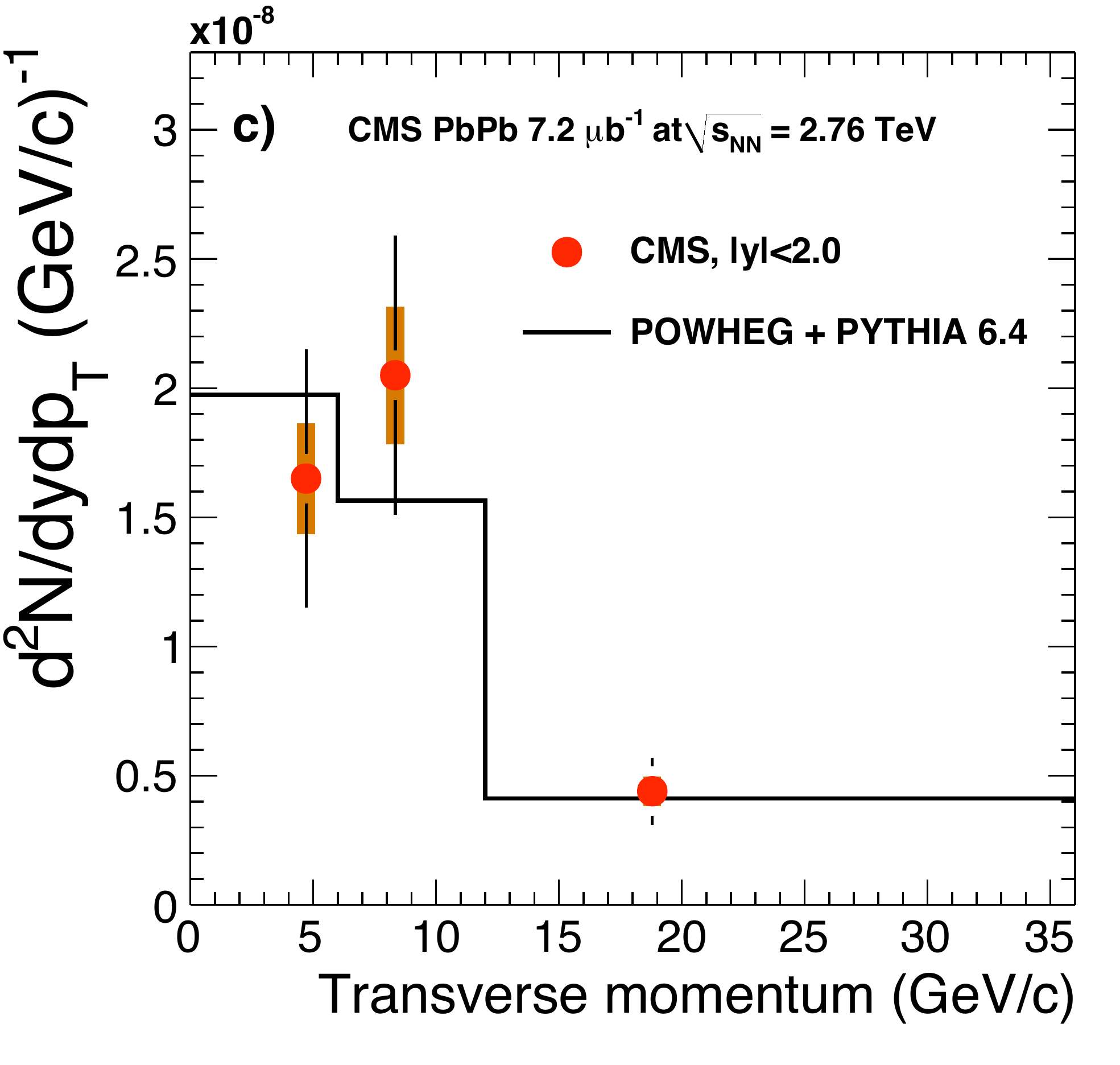}
    \caption{The yields of $\cPZ \rightarrow \mu\mu$ per event: a) $dN/dy$ divided by the expected nuclear overlap function $T_{\rm AB}$ and as a function of event centrality parameterised as the number of participating nucleons $N_{\rm part}$, b) $dN/dy$ versus the \cPZ\ boson $y$, c) $d^2N/dydp_{\rm T}$ versus the \cPZ\ boson $p_{\rm T}$. Data points are located horizontally at average values measured within a given bin. Vertical lines (bands) correspond to statistical (systematic) uncertainties. Theoretical predictions are computed within the same bins as the data, and are described in the text.}
    \label{fig:yields}
  \end{center}
  \vspace{-2ex}
\end{figure}

The full circles in Fig.~\ref{fig:yields} (a) show the centrality dependence of the \cPZ\ yield divided by $T_{\rm AB}$, while the open square is for MB events. The variable used on the abscissa is the average number of participating nucleons $N_{\rm part}$ corresponding to the selected centrality intervals, computed in the same Glauber model. No centrality dependence of the binary-scaled \cPZ\ yields is observed in data. A similar result was recently published by the ATLAS collaboration~\cite{Atlas:2010px}.

The normalised yields $(dN/dy)/T_{\rm AB}$ are compared to various calculations: 1) using the nucleon CT10 and modified nuclear EPS09 PDFs~\cite{Eskola:2009uj,Paukkunen:2010qg}, 2) using MSTW08 PDFs~\cite{Martin:2009iq} and modelling energy loss~\cite{Neufeld:2010fj}, and 3) provided by the \textsc{powheg}~\cite{Alioli:2008gx} generator interfaced with the \PYTHIA parton-shower generator and using CTEQ6.6 PDFs~\cite{Pumplin:2002vw}. Only a marginal centrality dependence is predicted: the inhomogeneous (i.e.~depending on the radial position in nuclei) shadowing is predicted to have negligible impact~\cite{Vogt:2000hp} and the energy-loss prediction drops by 3\% from peripheral to central collisions~\cite{Neufeld:2010fj}.

Figure~\ref{fig:yields} (b) and (c) show the differential yields, $dN/dy$ and $d^2N/dydp_{\rm T}$, as a function of the \cPZ\ boson $y$ and $p_{\rm T}$. They are compared to the same theoretical calculations as used for the centrality distribution (when available) multiplied by the minimum bias $T_{\rm AB}$ value. No significant deviations from binary-collision scaling are observed.

Nuclear modification factors, $R_{\rm AA}=dN / (T_{\rm AB} \times d\sigma_{\Pp\Pp})$, are computed from the AA measured yields $dN$, the nuclear overlap function $T_{\rm AB}$, and the $\Pp\Pp\rightarrow$~\cPZ\ cross sections $d\sigma_{\Pp\Pp}$ given by the \textsc{powheg} calculation (solid lines on Fig.~\ref{fig:yields}, e.g. $d\sigma_{\Pp\Pp}/dy = 59.6$~pb in $|y|<2.0$). The $R_{\rm AA}$ systematic uncertainty includes $T_{\rm AB}$ uncertainties, but no uncertainty is assigned to the theoretical \Pp\Pp\ cross section. All $R_{\rm AA}$ values are found compatible with unity. They are reported in Table~\ref{tab:stat}, together with the number of observed \cPZ\ bosons and their yield per event.

\begin{table}[hbtp]
\begin{center}
\vspace{-2ex}
\caption{For each $|y|$, $p_{\rm T}$ and centrality interval, number of \cPZ\ bosons $N_{\cPZ}$, associated yield per event $dN/dy$, and nuclear modification factor $R_{AA}$ derived by using a \textsc{powheg} \Pp\Pp\ reference. The quantity $d^2N/dy dp_{\rm T}$ is given in units of (GeV/$c)^{-1}$. The first uncertainty is statistical and the second systematic.\label{tab:stat}}
\vspace{1ex}
\begin{tabular}{|c|c|c|c|}
  \hline
  $|y|$ &      $N_{\cPZ}$ & $dN/dy$ ($10^{-8}$)  & $R_{\rm AA}$ \\ \hline \hline
  $[0,2.0]$ &     39 & $33.8 \pm 5.5  \pm 4.4 $ & $1.00 \pm 0.16 \pm 0.14$ \\ \hline
  $[0,0.5]$  &    13 & $38.1 \pm 10.7 \pm 5.0 $ & $1.03 \pm 0.29 \pm 0.15$ \\
  $[0.5,1.0]$ &   12 & $35.6 \pm 10.4 \pm 4.6 $ & $0.98 \pm 0.29 \pm 0.14$ \\
  $[1.0,2.0]$  &  14 & $30.0 \pm 8.1  \pm 3.9 $ & $0.97 \pm 0.26 \pm 0.14$ \\ \hline \hline
  $p_{\rm T} (GeV/$c$)$ & $N_{\cPZ}$ & $d^2N/dydp_{\rm T}$ ($10^{-8}$)  & $R_{AA}$ \\ \hline
  $[0,6]$  &  11 & $1.65 \pm 0.50 \pm 0.22 $ & $0.84 \pm 0.26 \pm 0.12$ \\
  $[6,12]$ &  15 & $2.05 \pm 0.54 \pm 0.27 $ & $1.32 \pm 0.34 \pm 0.19$ \\
  $[12,36]$ & 12 & $0.44 \pm 0.13 \pm 0.06 $ & $1.06 \pm 0.31 \pm 0.15$ \\ \hline \hline
  Centrality &      $N_{\cPZ}$ & $dN/dy$ ($10^{-8}$)  & $R_{AA}$ \\ \hline
  $[30,100]$\% & 7 & $7.9 \pm 3.0 \pm 1.0 $ & $0.92 \pm 0.35 \pm 0.16$ \\
  $[10,30]$\% & 14 & $59.5 \pm 16.0 \pm 7.7  $ & $0.86 \pm 0.23 \pm 0.12$ \\
  $[0,10]$\% &  18 & $165  \pm   40 \pm 22   $ & $1.20 \pm 0.29 \pm 0.16$ \\
    \hline
\end{tabular}
\end{center}
  \vspace{-2ex}
\end{table}

In conclusion, the \cPZ\ boson yield in PbPb collisions at $\sqrt{s_{\rm NN}} = 2.76$~TeV has been measured inclusively and as a function of rapidity, transverse momentum, and centrality. Within uncertainties, no modification is observed with respect to theoretical next-to-leading order pQCD proton-proton cross sections scaled by the number of elementary nucleon-nucleon collisions. This measurement confirms the validity of the Glauber scaling for perturbative cross sections in nucleus-nucleus collisions at the LHC and establishes the feasibility of carrying out detailed \cPZ\ physics studies in heavy-ion collisions with the CMS detector. With upcoming PbPb collisions at higher luminosity, the \cPZ\ boson promises to be a powerful reference tool for final-state heavy-ion related signatures as well as providing a means to study the modifications of the parton distribution functions.

We thank Bryon Neufeld, Hannu Paukkunen, Carlos Salgado, Ivan Vitev, and Ramona Vogt for fruitful theoretical inputs on the  nuclear effects involved in \cPZ\ production.
We wish to congratulate our colleagues in the CERN accelerator departments for the excellent performance of the LHC machine in 2010. We thank the technical and administrative staff at CERN and other CMS institutes, and acknowledge support from: FMSR (Austria); FNRS and FWO (Belgium); CNPq, CAPES, FAPERJ, and FAPESP (Brazil); MES (Bulgaria); CERN; CAS, MoST, and NSFC (China); COLCIENCIAS (Colombia); MSES (Croatia); RPF (Cyprus); Academy of Sciences and NICPB (Estonia); Academy of Finland, ME, and HIP (Finland); CEA and CNRS/IN2P3 (France); BMBF, DFG, and HGF (Germany); GSRT (Greece); OTKA and NKTH (Hungary); DAE and DST (India); IPM (Iran); SFI (Ireland); INFN (Italy); NRF and WCU (Korea); LAS (Lithuania); CINVESTAV, CONACYT, SEP, and UASLP-FAI (Mexico); PAEC (Pakistan); SCSR (Poland); FCT (Portugal); JINR (Armenia, Belarus, Georgia, Ukraine, Uzbekistan); MST and MAE (Russia); MSTD (Serbia); MICINN and CPAN (Spain); Swiss Funding Agencies (Switzerland); NSC (Taipei); TUBITAK and TAEK (Turkey); STFC (United Kingdom); DOE and NSF (USA).

\bibliography{auto_generated}   
\cleardoublepage\appendix\section{The CMS Collaboration \label{app:collab}}\begin{sloppypar}\hyphenpenalty=5000\widowpenalty=500\clubpenalty=5000\textbf{Yerevan Physics Institute,  Yerevan,  Armenia}\\*[0pt]
S.~Chatrchyan, V.~Khachatryan, A.M.~Sirunyan, A.~Tumasyan
\vskip\cmsinstskip
\textbf{Institut f\"{u}r Hochenergiephysik der OeAW,  Wien,  Austria}\\*[0pt]
W.~Adam, T.~Bergauer, M.~Dragicevic, J.~Er\"{o}, C.~Fabjan, M.~Friedl, R.~Fr\"{u}hwirth, V.M.~Ghete, J.~Hammer\cmsAuthorMark{1}, S.~H\"{a}nsel, C.~Hartl, M.~Hoch, N.~H\"{o}rmann, J.~Hrubec, M.~Jeitler, G.~Kasieczka, W.~Kiesenhofer, M.~Krammer, D.~Liko, I.~Mikulec, M.~Pernicka, H.~Rohringer, R.~Sch\"{o}fbeck, J.~Strauss, F.~Teischinger, P.~Wagner, W.~Waltenberger, G.~Walzel, E.~Widl, C.-E.~Wulz
% ...
\vskip\cmsinstskip
% ...
1:~~Also at CERN, European Organization for Nuclear Research, Geneva, Switzerland\\
% ...

\end{sloppypar}
\end{document}